\documentclass[twocolumn]{article}

\usepackage{scrextend}
\usepackage{lineno,hyperref}
\usepackage{subfig}
\usepackage{graphicx}
\usepackage{amsmath}
\usepackage[affil-it]{authblk}
\begin{document}

\title{\vspace{-2.0cm}\textbf{Genetic evolution of a multi-generational population in the context of interstellar space travels -- Part II: Phenotypic effects of gene expression}}
   
\author{Fr\'ed\'eric Marin\textsuperscript{1}, Camille Beluffi-Marin\textsuperscript{2} \& Fr\'ed\'eric Fischer\textsuperscript{3}\\
{\small 1 Universit\'e de Strasbourg, CNRS, Observatoire astronomique de Strasbourg, UMR 7550, F-67000 Strasbourg, France\\
2 INSERM, UMRS\_1112, Strasbourg, France\\
3 Institut de physiologie et de chimie biologique, Laboratoire Dynamique \& Plasticit\'e des Synth\'etases, UMR 7156, F-67000 Strasbourg, France}}
              
\date{Dated: \today}

\twocolumn[
  \begin{@twocolumnfalse}
    \maketitle
    \begin{abstract}
In the first paper of this series, we included the effects of population genetics in the agent-based Monte Carlo
code HERITAGE under the hypothesis of neutral phenotypic effects. It implied that mutations (genetic changes) 
had only neutral physical manifestations. We now relax this assumption by including genetic effects of mutation 
and neo-mutations (from radiations) onto the population's life expectancy, fertility, pregnancy chances and 
miscarriage rates. When applied to a population aboard a generation ship that travels at sub-light speed towards 
a distant exoplanet, we demonstrate that natural selection indirectly affects the genetic structure of a population 
via the contribution of phenotypes, in  agreement with past studies in conservation biology.  
For large starting crews (about 500 individuals), the effect aligns with the neutral hypothesis and the frequency 
of alleles (for non-sexual chromosomes) is stable over centuries. Results are completely different if the spacecraft 
shielding, integrated into hull design, fails to efficiently protect the crew from high-energy cosmic rays and showers
of secondary particles. We tested different scenarios, in which the level of radiation is either fixed at normal 
or extreme levels, or changing over time due to, e.g., shield degradation, on-board nuclear incident or the outburst 
of a supernova situated 50 light-years away.
    \end{abstract}
    
    \vspace{3\baselineskip}
  \end{@twocolumnfalse}
]

%%%%%%%%%%%%%%%%%%%%%%%%%%%%%%%%%%%%%%%%%%%%%%%%%%%%%%%%%%%%%%%%%%%%%%%%%%%%
%%%%%%%%%%%%%%%%%%%%%%%%%%%%%%%%%%%%%%%%%%%%%%%%%%%%%%%%%%%%%%%%%%%%%%%%%%%%
\section{Introduction}
\label{Introduction}
As humanity ventures further into the exploration of space, the prospect of interstellar travel presents both unprecedented opportunities and formidable challenges \cite{Neukart2024}. Among the difficulties facing long-duration space missions, the biological adaptation of organisms to the unique environmental conditions of deep space stands representing a critical concern. In particular, the genetic evolution of multi-generational populations subjected to the rigors of interstellar travel represents a complex frontier of scientific inquiry \cite{Marin2021}. Indeed, interstellar space travel introduces a multitude of stressors and environmental factors distinct from those experienced on Earth or within our immediate cosmic vicinity. Exposure to extreme cosmic radiation \cite{Chancellor2014}, prolonged microgravity \cite{Bradbury2020}, limited resource availability \cite{Marin2020}, and isolation from terrestrial ecosystems \cite{Kanas2015} represent some of the primary challenges faced by organisms embarking on interstellar journeys. In the face of such adversity, the ability of organisms to adapt and evolve becomes critical for ensuring the success and sustainability of long-duration space missions.

Genetic evolution, the process by which heritable traits are passed from one generation to the next and subjected to natural selection, offers a mechanism through which organisms can potentially adapt to the novel conditions encountered during interstellar travel. While much attention has been devoted to the physiological and psychological effects of space travel on astronauts \cite{Chancellor2014}, relatively little is known about the genetic mechanisms underlying long-term adaptation to space environments, especially over multiple generations \cite{Smith2014} and under stronger radiation doses than what is currently experienced on Earth \cite{NRC1990}. Understanding how gene expression influences phenotypic traits is essential for predicting how organisms will adapt to the unique challenges of space travel. Phenotypic traits, such as resistance to radiation, metabolic efficiency and immune function, directly impact an organism's ability to survive and thrive in space environments. Incorporating phenotypic effects into genetic evolution simulations enables us to better predict the adaptive responses of populations over multiple generations in a space-limited environment.

For example, recent studies have shown that microgravity can significantly alter gene expression. Research coordinated by the European Space Agency (ESA) and conducted by the University of Surrey found that prolonged exposure to simulated microgravity (via a 60-day bed rest protocol) disrupted rhythmic gene expression in humans. Approximately 91\% of the genes studied exhibited significant alterations in their number, timing, and amplitude of expression \cite{Archer2024}. It illustrates the fact that the emergence of rare diseases is favored in the context of deep space travels, since these diseases are 80\% genetic in origin and result from mutations in specific genes, which can be inherited or occur de novo (new mutations). These mutations can affect protein function, gene regulation, or other cellular aspects, leading to varied clinical conditions \cite{Cooper1992}. Rare diseases exhibit great clinical heterogeneity, making their identification difficult. They may manifest symptoms common to other more common diseases, leading to misdiagnosis or delays. In addition, the same genetic mutation may manifest with very different symptoms from one patient to another, even within the same family. This phenotypic variability complicates the identification of specific causative mutations and makes it difficult to establish an accurate diagnosis based on clinical symptoms alone \cite{Girirajan2010}. Although next-generation sequencing (NGS) technologies have revolutionized genetic diagnosis, they also have their limitations. Some mutations, such as large deletions or duplications, may not be effectively detected by these technologies. In addition, non-coding variants (regions of the genome that do not code for proteins) may play an important role in some rare diseases, but they are often poorly understood and difficult to interpret \cite{Pagnamenta2023}. Thus, at the present time and with the state-of-the-art of our knowledge in genetics, sending a crew into space for a long period subjects it to an increased risk of seeing its members develop rare diseases for which no hindsight will be possible as to the impact of these on these members and ultimately on the mission, hence the need for numerical simulations such as presented in this paper.

In this second paper of our series, we aim to examine the genetic evolution of multi-generational populations during interstellar space travel, with a specific focus on the phenotypic effects of gene expression. To do so, we will use the agent-based Monte Carlo code HERITAGE developed by our team  \cite{Marin2017,Marin2018,Marin2019,Marin2020,Marin2021}, which now includes phenotypic effects of gene expression on several key quantitative biological (likelihood of infertility, number of miscarriage, pregnancy chances) and genetical indicators (genome diversities with respect to the reference genome, genetic polymorphism, degree of genetic heterozygosity and Nei’s genetic distance). The methods for modeling environmental and radiation-induced phenotypic effects is presented in Sect.~\ref{Modeling}. Five simulations representative of the various radiation exposure scenarios that a multi-generational vessel could encounter are presented in Sect.~\ref{Results} and are analyzed in Sect.~\ref{Analysis}. We conclude our paper and envision the next steps for HERITAGE in in Sect.~\ref{Conclusions}.

%%%%%%%%%%%%%%%%%%%%%%%%%%%%%%%%%%%%%%%%%%%%%%%%%%%%%%%%%%%%%%%%%%%%%%%%%%%%
%%%%%%%%%%%%%%%%%%%%%%%%%%%%%%%%%%%%%%%%%%%%%%%%%%%%%%%%%%%%%%%%%%%%%%%%%%%%
\section{Modeling phenotypic effects}
\label{Modeling}
In this section, we present the improvements we made in our Monte Carlo, agent-based code HERITAGE for simulating
realistic human population in resource-restricted, enclosed environments. Those upgrades are centered around relaxing 
the neutral selection hypothesis on genetic evolution for a multi-generational population in the context of interstellar
space travels.

\subsection{Brief reminder on genetics modeling in HERITAGE}
\label{Modeling:reminder}

In HERITAGE \cite{Marin2021}, we model simplified diploid human genomes using matrices. Each matrix represents a DNA segment, 
i.e. a chromosome, and each numerical human has 46 physically independent chromosomes (23 of paternal origin and 23 of maternal 
origin). Along each chromosome, we define discrete and bounded blocks that are independent of each other in the sense that we 
can distinguish them by a property. Those are the loci (singular ``locus''). Each locus can have 10 various allelic forms, representing
the genetic variations that exist between individuals and between human populations. In total, there are 2\,110 loci for a single 
entire diploid genome in HERITAGE, with 21\,100 different possible allelic forms.

Of course, not all humans have the same genome (they are not clones). At the beginning of the simulation, we first define a standard reference 
human genotype by setting all alleles to 0. Then, we create pools of 100 individual genotypes with a specific variation level 
and cross these 100 genotypes in a randomized fashion to create five final populations. Our final crew consists of a user-defined 
number of randomly selected individuals among these 5 reference populations. As stated in \cite{Marin2021}, this ``makes it possible 
to account for the fact that these populations, even if they are the initial ones, are themselves the result of a complex 
(and common) genetic history, with varying levels of genetic relatedness''.

Because each crew member of the zeroth-generation has a specific genotype, we are able to create realistic new generations of digital 
humans following the rules of heredity. Meiosis (the process of double-cell division that allows switching from a diploid cell to 
four haploid cells), duplication, homologous recombination, conversion, separation and distribution are included. The process is the same for 
egg formation and sperm formation, so we do these genetic tasks for both the mother and the father to obtain a genetically plausible 
offspring. Using this methodology, each novel individual possesses a unique genotype inherited from his/her parents. 

All this work was achieved under the neutral hypothesis condition, i.e. no phenotypic effect of mutations. Consequently, there was no
natural selection of alleles or allelic patterns, leading to little-to-no genetic differentiation of the n$^{\rm th}$ population with respect to 
the starting one. This is the hypothesis we will now relax.

\subsection{Phenotypic effects of mutations (natural selection)}
\label{Modeling:natural}

Let us consider our initial population from a genetic point of view. It is a lasting population that has been able to survive and 
reproduce on Earth for thousands of years, meaning that the mutations in their genome are mostly non-deleterious. Any strongly disadvantageous 
mutations were purged from the population over the centuries, since they lead to the premature death (or infertility) of the carriers 
\cite{Gillespie2004}. It means that the existing allelic combinations in our population have essentially a neutral effect, but problems 
might occur when new allelic combinations appear.

To take this effect into account, at the beginning of each new HERITAGE simulation, we scan the genome of each initial crew member and save 
all the existing allelic combinations. These combinations and their respective phenotypic effects are considered as neutral. To express 
this neutrality, a weight is applied to each allelic combination. For a neutral phenotypic effect, this weight is set to unity. However, 
for all new combinations within the 69\,630 possible ones that do not exist in the initial population, non-unitarian weights 
are stochastically associated. We save each of the weighted allelic combinations in a file so that the phenotypical effects will stay the same 
at any time during the simulation. 

This weight is the genetic fitness, i.e. the capability of an organism (here a specific allele) to 
survive and reproduce. The phenotypic expression of the genotype in a particular environment determines how genetically fit an organism will 
be. If the fitness value is lower than unity, the allelic combination is deleterious. If the weight is larger than 1, the mutation is beneficial. 
In HERITAGE, the fitness is applied to 4 main parameters: the digital humans' life expectancy, fertility, pregnancy chances and miscarriage 
rates. More phenotypic effects can be included but, for this paper, we will focus on the main ones that drive the population survival chances 
over long periods.

The distribution of fitnesses for natural mutations (not induced by external irradiation) is not unique. The simplest, yet rather efficient, 
models use a Gaussian distribution centered around 1 and with a constant width across environments \cite{Burger2002,Martin2006}. Of course, 
white noise (random processes uncorrelated in both space and time) can be added to the Gaussian distribution, that can also be multivariate, 
to mimic life's stochasticity. Polynomial functions could also be used. In fact, there is a whole landscape for the distribution of fitness
effects \cite{Tenaillon2007,Bataillon2014}. In HERITAGE, the distribution of fitness effects follow the most-often used Gaussian function. 
The height of the curve's peak is set to a normalized value of 1, the position of the center of the fitness peak is fixed to 1, and the 
standard deviation is at the user's choice (fixed to 0.025 by default).

During the simulation, every time there is a successful procreation event, the genome of the offspring is scanned. All allelic combinations are
compared to their associated fitness value and the mean genetic fitness for each phenotypic effect is computed. Because selection over many 
generations is a multiplicative process, we multiply the various fitnesses together to obtain a representative genetic fitness for each of the four
effects studied in this paper (life expectancy, fertility, pregnancy chances or miscarriage rates). The final four genetic fitness values are 
then multiplied to their associated biological data quoted just before (each their own). In the case of deleterious combinations, the life expectancy 
(or any other three parameters) of the person will decrease, resulting in the natural selection of the best allelic combinations. As a consequence, 
natural selection does not have an immediate effect, which would automatically replace an allele with another more advantageous, or eliminate 
alleles with a deleterious effect. Rather, it introduces an additional parameter for the probability of fixation of alleles with respect to 
each other. This parameter accounts for the fact that carriers of one of the two (or more) competing alleles will have a better probability 
of survival and/or better reproductive capacities, and will therefore leave more offspring on average. The greater the difference between 
the fitness of the two competing alleles, the stronger the effect will be.

\subsection{Phenotypic effects of neo-mutations (irradiation)}
\label{Modeling:irradiation}

\begin{figure}
\centering
\includegraphics[trim = 0mm 0mm 0mm 0mm, clip, width=8cm]{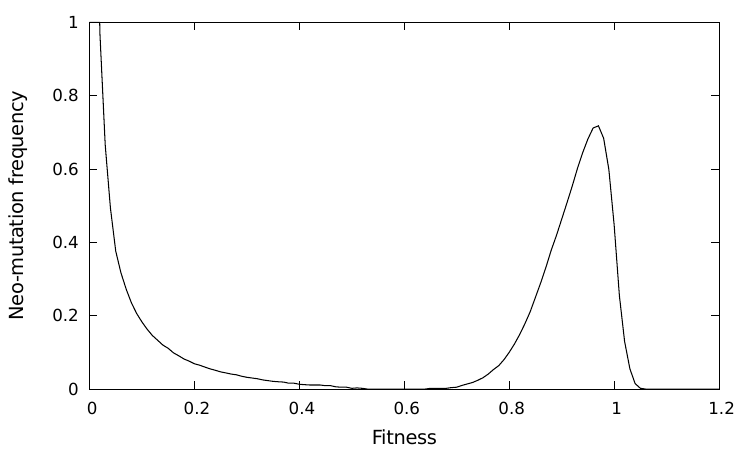}
  \caption{Bimodal distribution of fitness effects of neo-mutations \cite{Masel2013}.}
  \label{Fig:Distribution}
\end{figure}

A second source of genetic mutations comes from cosmic irradiation. While the Earth's atmosphere is opaque to cosmic rays energies below 1~GeV, 
the flux of high-energy galactic cosmic rays (along with radiation from solar proton events and the radiation belts) greatly increases with 
increasing distance from our planet. The risks for biological damage from primary galactic cosmic rays bombardment or showers of secondary 
particles emerging from the interaction of cosmic rays with the spacecraft shields are high in deep space. Ionizing radiation can cause 
different types of DNA lesions, base modifications, DNA-protein bridging, or even strand breaks. When the lesions can be repaired, the biological 
effect is then limited to the molecular level and the integrity of the cell is preserved. When the cell cannot repair its damaged DNA, 
the modifications cause prolonged arrests in the cell cycle, inhibition of transcription and DNA replication, and chromosome segregation. 
May then appear mutations and chromosomal aberrations. Radio-biological effects on the reproductive system imply infertility, miscarriage, 
decrease of pregnancy chances, and/or malformations \cite{Ogilvy1993} as well as deadly cancers \cite{Sachs2005,Imaoka2009}. Those parameters
are the one treated by HERITAGE.

To model neo-mutations, i.e. mutations caused by external irradiation, the code now accounts for a 11$^{\rm th}$ allelic state. Any mutation 
induced by radiation may change the current allelic state of a locus to this peculiar value, highlighting the fact that a new (hereditary) 
mutation appeared in the genotype. At first approximation, the number of neo-mutations that the genome of a given crew member may face per year 
is directly proportional to the irradiation dose at low energies ($\le$ 1~Sv) and becomes exponential at higher energies. The lower part is a 
standard linear no-threshold model, meaning that an increase in dose results in a proportional increase in risk and any dose, no matter how small, 
produces some genetic risk \cite{Radiation1996}. The high energy part of the formula is upwardly concave, i.e. exponential: the mutation rate
is doubled per 1~Sv absorbed \cite{Roberts1976,Cucinotta1997,Nakamura2013}. This can be modeled using the following equation:

\[ N_i = exp(-13.5924 + 0.6931 \times R_i) \times 2 \times G \]

with $N_i$ the number of neo-mutations appearing per year in the genome of the $i^{\rm th}$ crew member, $R_i$ the accumulated radiation dose
of the $i^{\rm th}$ crew member (in Sv), the factor 2 is the number of sets of each gene and $G$ is the total number of genes per individual
(54\,083). The numerical values of the exponential are derived from the formula for the number of spontaneous new mutants per generation given 
by \cite{BEIR1972,Roberts1976}:  $G \times N \times 2 \times m$, with $N$ the number of individuals (1) and $m$ the spontaneous mutation rate per gene
per generation (1.25 $\times$ 10$^{-6}$). We checked that our resulting formula gives the expected number of mutational events per genome in 
protein-coding genes after 1~Sv exposure: 0.27 \cite{Nakamura2013}. 

Because $R_i$ is time-dependent, the effect of various radiation doses can be investigated throughout the interstellar journey. However, the 
distribution of fitness effects for neo-mutations does not follow a Gaussian function. It is rather bimodal, with an asymmetric bell-curve 
centered below 1 (beneficial effects are rarer than deleterious effects), and a tail of lethal fitness effects \cite{Masel2013}. This specific
distribution is shown in Fig.~\ref{Fig:Distribution} but it can be changed according to the user needs. This function tells us that neo-mutations
are predominantly detrimental to the human body and high radiation doses could potentially decimate entire populations.

\subsection{The chromosome map}
\label{Modeling:map}

In order to investigate the phenotypic effects of mutations and neo-mutations, HERITAGE now takes as an input a ``chromosome map''.
This map is a list of all loci that can show a gene expression, the other (non-listed) ones being considered as neutral with respect
to the four mains phenotypic effects included in the code: life expectancy, fertility, pregnancy chances and miscarriage rates. 
The user can select any number of locus positions on any number of chromosomes and impose an effect on one, two, three or the four 
gene expressions defined above. A typical entry reads as:

\[ Ch \cdot l \cdot E_{\rm fert} \cdot E_{\rm life} \cdot E_{\rm preg} \cdot E_{\rm miss} \]

with $Ch$ the chromosome number, $l$ the locus number and $E_{\rm X}$ a yes-or-no string that indicates if the locus has a phenotypic 
manifestation. $X$ represents, in this order, fertility, life expectancy, pregnancy chances and miscarriage rate. If one or several 
$E_{\rm X}$ have a ``yes'' value, all allelic combinations (0,0; 0,1; 0,2 ... 9,10; 10,10) for the $X$ parameter are weighted by a
fitness. Allelic values between 0 and 9 follow the distribution of fitness effects for natural mutations (a Gaussian in our case, as 
explained in Sect.~\ref{Modeling:natural}) and all allelic values equal to 10 are neo-mutations whose fitness values follow the bimodal 
distribution shown in Fig.~\ref{Fig:Distribution} and detailed in Sect.~\ref{Modeling:irradiation}. For each allelic combination, the 
dominant and recessive alleles are randomly selected and fixed for the rest of the simulation. We remind that dominance refers to the 
relationship between the two alleles of a gene. If they are different, one allele will be expressed; it is the dominant gene. The effect 
of the other allele, called recessive, is ignored.

%%%%%%%%%%%%%%%%%%%%%%%%%%%%%%%%%%%%%%%%%%%%%%%%%%%%%%%%%%%%%%%%%%%%%%%%%%%%
%%%%%%%%%%%%%%%%%%%%%%%%%%%%%%%%%%%%%%%%%%%%%%%%%%%%%%%%%%%%%%%%%%%%%%%%%%%%
\section{Genetic effects over 600 years of interstellar travel}
\label{Results}
To investigate the effects of gene mutations, we now simulate a 600 year-long interstellar travel towards a potential exoworld.
Knowing that the speed growth of artificial vehicles is about 4.72\% annually (doubling every 15 years \cite{Heller2017}),
such mission could very well be directed towards the closest exoplanetary system by the next century.

\subsection{Input parameters}
\label{Results:model}
The generation ship capacity is set to 1\,200 persons, with a 90\% overpopulation threshold (procreation is no longer allowed at 90\% 
of the ship total carrying capacity). Our baseline model for the initial population comprises a crew of 500 persons, equally distributed 
in gender. On average, the departing crew members are 30 years old, with a Gaussian standard deviation of 5 years. The allowed procreation 
window is set to 18 -- 40 years old and the age of menopause is fixed to 45. Each woman can have 3 children with a standard deviation of 1 
(to be modulated by social restrictions when the overpopulation threshold will be reached). The women's life expectancy is set to 85 years 
with a standard deviation of 5, while men's life expectancy is set to 79 years with the same standard deviation. The initial crew members 
do not show consanguinity between each others and reproduction is allowed up to a consanguinity factor of 3\%. Those parameters are 
very common for generation ships initial crews \cite{Marin2017,Marin2018,Marin2019,Marin2020,Marin2021} and their different impacts have 
be thoroughly tested in the first papers of this series.

In terms of population genetic, the departing crew members are hand-picked from genetically realistic initial populations that share a 
common history but do not show strongly deleterious mutations. The typical difference between an individual's genome and the reference 
genome was fixed to 0.6\% of the total of human basepairs \cite{Auton2015}. We take into account the radiation dose received by the 
crew on Earth before departure on the assumption that the members were not subjected to abnormal doses of ionizing radiation. Our
chromosome map was built so that 3.5\% of the whole locus pool has a phenotypic effect onto the life expectancy, fertility, pregnancy 
chances and miscarriage rate of the population. The true percentage is unknown since there is a huge number of candidate genes to be 
studied in laboratory \cite{Zorrilla2013}. However, we chose this number to replicate the widely used figure of a 5\% excess risk of 
death from cancer with an accumulated radiation dose of 1~Sv \cite{ICRP1991,NCRPM1993}. Conveniently, our parametrization also reproduce 
the expected 50\% / 100\% causality rate from acute radiation doses larger than 5 and 10~Sv, respectively \cite{Anno2003}. 

For this set of simulations, we do not implement diseases or catastrophes in order to focus on the phenotypic effects of mutations. 
In the first paper of this series, we have already seen that catastrophes create population bottlenecks, where episodes of severe population 
reduction followed by further population expansions lead to a reduction in the genetic diversity of the crew \cite{Marin2021}.

\subsection{Scenario 1: Earth-like natural background radiation}
\label{Results:Earth}
\begin{figure*}
\centering
  \subfloat[Evolution of the population within the space ship.]{\includegraphics[width=8cm]{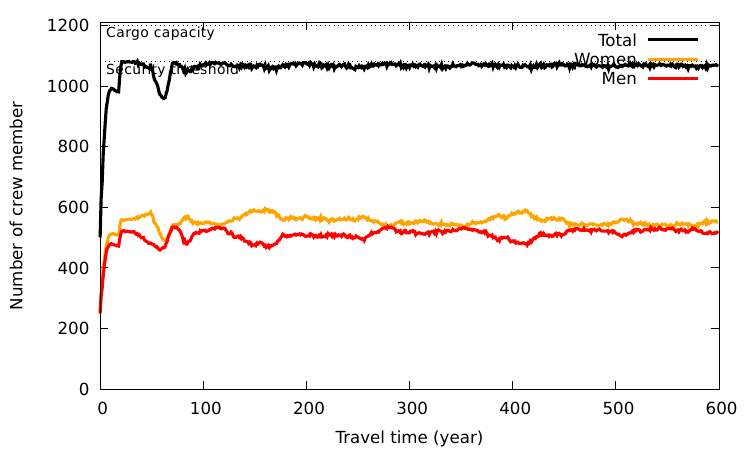}}\hfil
  \subfloat[Equivalent dose of ionizing radiation (in mSv).]{\includegraphics[width=8cm]{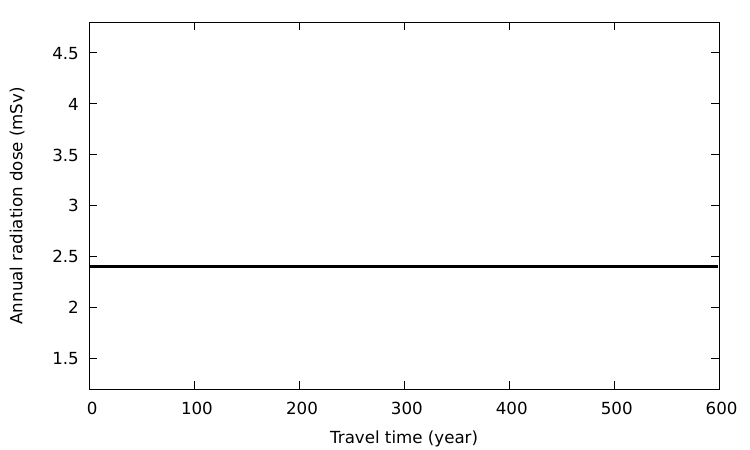}}

  \subfloat[Likelihood of infertility among breeding people.]{\includegraphics[width=8cm]{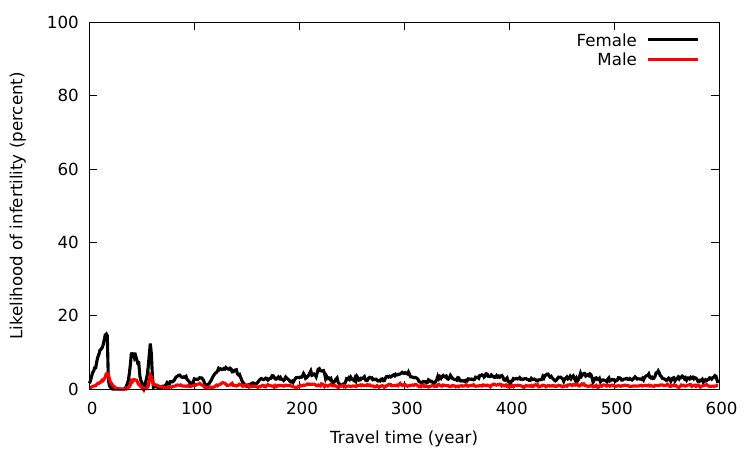}}\hfil
  \subfloat[Likelihood of pregnancy among breeding women.]{\includegraphics[width=8cm]{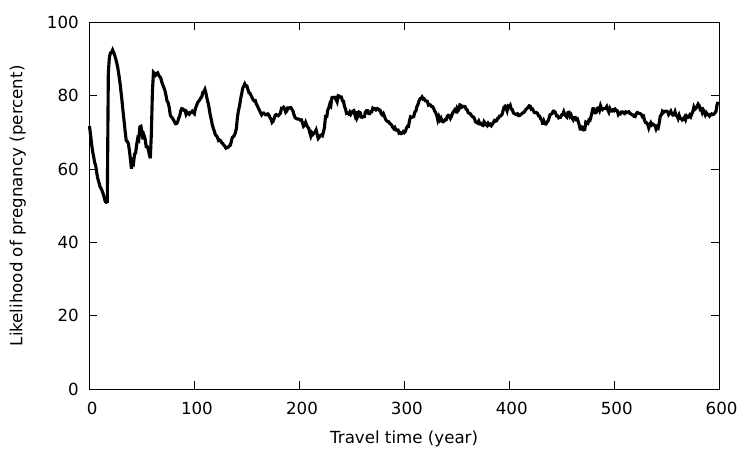}} 

  \subfloat[Number of miscarriages.]{\includegraphics[width=8cm]{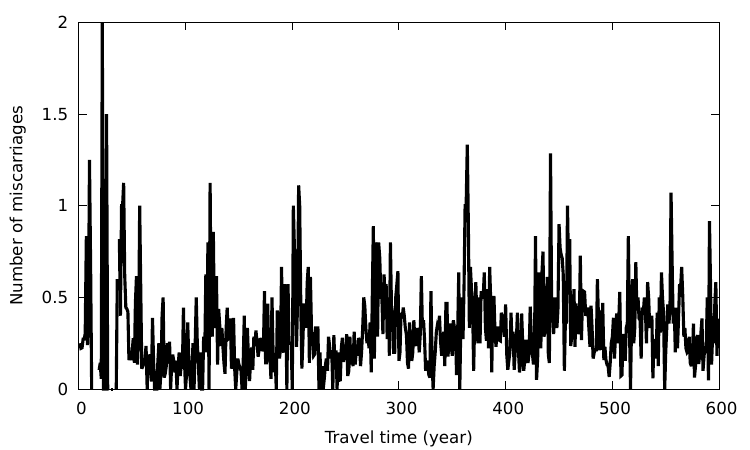}}\hfil   
  \subfloat[Individual heterozygosity among the crew.]{\includegraphics[width=8cm]{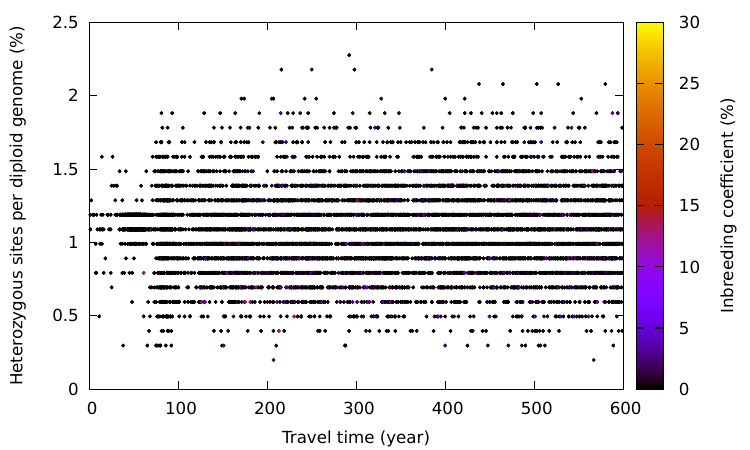}}
  \caption{Results from the first scenario (Earth-like natural background radiation).}
  \label{Fig:Scenario1}
\end{figure*}

Our first model is built under the hypothesis of a perfect radiation shield that prevents any permeation of cosmic rays from the interstellar medium 
and no nuclear incident inside the generation ship over the course of the spacecraft. The radiation dose received by the crew members is constant 
in time and fixed to 2.4~mSv per year \cite{UNSC2010}, a value representative of Earth-like natural background radiation outside sites and soils 
polluted by radioactive substances. The results of our simulations are shown in Fig.~\ref{Fig:Scenario1} and will be analyzed in details in Sect.~\ref{Analysis}.

\subsection{Scenario 2: extreme background levels}
\label{Results:super-Earth}
\begin{figure*}
\centering
  \subfloat[Evolution of the population within the space ship.]{\includegraphics[width=8cm]{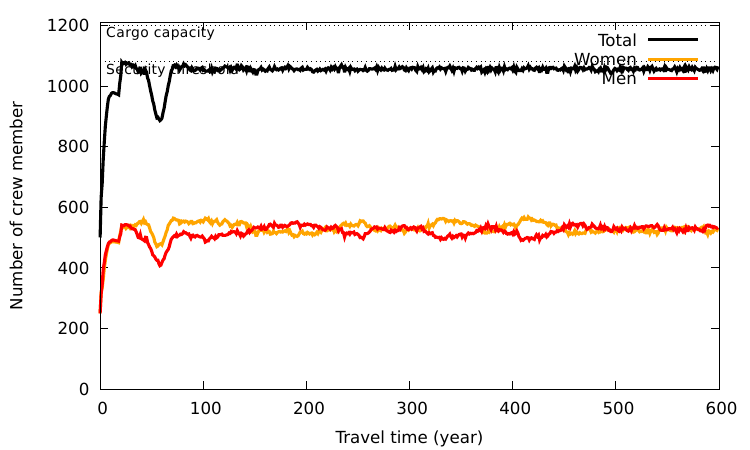}}\hfil
  \subfloat[Equivalent dose of ionizing radiation (in mSv).]{\includegraphics[width=8cm]{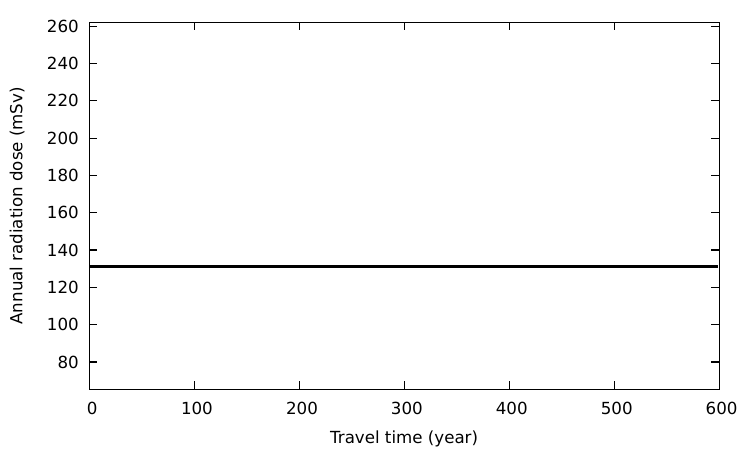}}

  \subfloat[Likelihood of infertility among breeding people.]{\includegraphics[width=8cm]{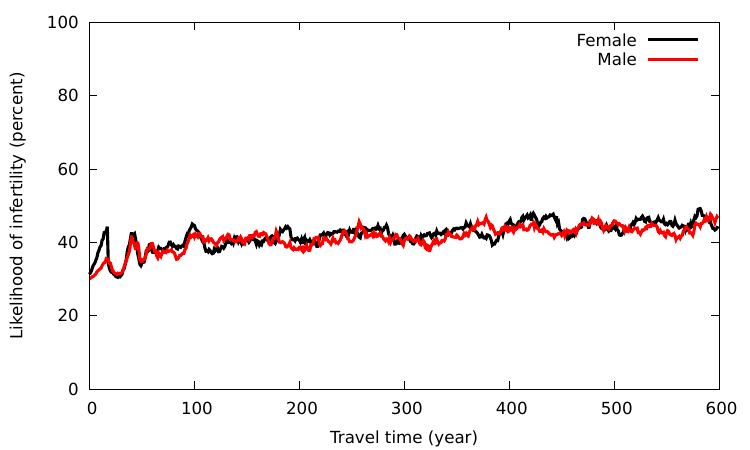}}\hfil
  \subfloat[Likelihood of pregnancy among breeding women.]{\includegraphics[width=8cm]{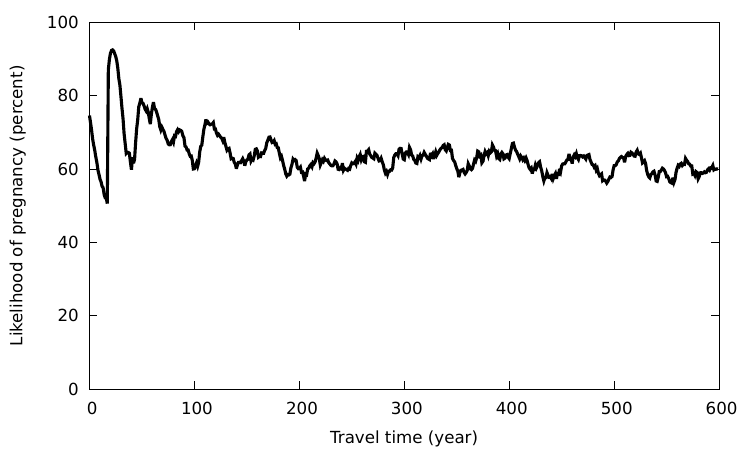}} 

  \subfloat[Number of miscarriages.]{\includegraphics[width=8cm]{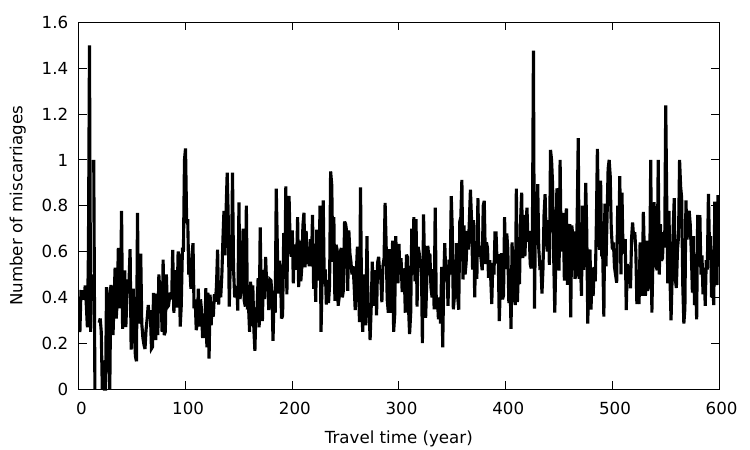}}\hfil   
  \subfloat[Individual heterozygosity among the crew.]{\includegraphics[width=8cm]{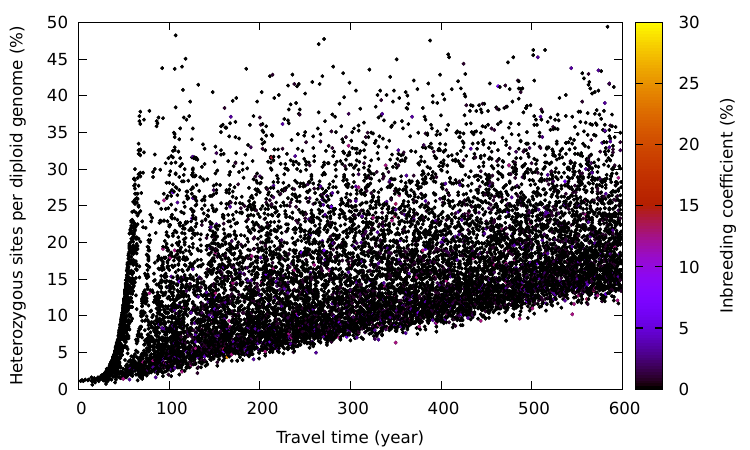}}
  \caption{Results from the second scenario (extreme background levels).}
  \label{Fig:Scenario2}
\end{figure*}

Our second model explores levels of radiation that can be considered as extreme cases of known inhabited regions on Earth. A perfect example
is the average radiation dose of 10~mSv per year received by the Ramsar's Talesh Mahalleh district (Iran), with the utmost case of 130~mSv per
year ambient field measured inside the most radioactive house \cite{Hendry2009}. This natural radioactivity originates from local geology, 
where underground water dissolves radium in uraniferous igneous rock and carries it to the surface through hot springs \cite{Ghiassi2002}. 
A fraction of the radium diffuses into the soil, where it mixes with drinking water. The rest of the radium precipitates into travertine, a form 
of limestone that has been used as a building material for decades. About 2\,000 people live in this district despite the high radiation doses. 
To investigate the long-term effects of such extreme conditions in a confined, space-based environment, we fixed the yearly radiation rate 
to a constant level of 130~mSv. The results of our simulations are shown in Fig.~\ref{Fig:Scenario2} and will be analyzed in details in Sect.~\ref{Analysis}.

%Early anecdotal evidence from local doctors and preliminary cytogenetic studies suggested that there may be no such harmful effect, and possibly even a radio-adaptive effect.[12] More recent epidemiological data show a slightly reduced lung cancer rate[13] and non-significantly elevated morbidity, but the small size of the population (only 1800 inhabitants in the high-background areas) will require a longer monitoring period to draw definitive conclusions.[14] Furthermore, there are questions regarding possible non-cancer effects of the radiation background. An Iranian study has shown that people in the area have a significantly higher expression of CD69 gene and also a higher incidence of stable and unstable chromosomal aberrations.[15] Chromosomal aberrations have been found in other studies[16] and a possible elevation of female infertility has been reported.[17]

\subsection{Scenario 3: progressive degradation of the radiation shield}
\label{Results:degradation}
Our third scenario postulates that the on-board radiation dose is at a safe level of 2.4~mSv per year but the spaceship shields are no
longer perfect. They progressively deteriorate with time. We impose a 0.01\% linear degradation of the protective layers per year, 
meaning that more and more interstellar radiation will permeate the shields over time. 

However, to know how much radiation permeates, it is necessary to first determine how much radiation can be expected from deep space. 
The Cosmic Ray Subsystem (CRS) instrument aboard Voyager-1, the space probe launched by NASA on September 5, 1977, allows us to make 
an educated guess. Considering the counting rate of galactic cosmic rays obtained from the High Energy Telescope 2 (HET 2) on the CRS 
from 1977 through 2015 \cite{Cummings2016}, one can see that the density of galactic cosmic nuclei and electrons is increasing until it 
reaches a plateau that corresponds to the entry of Voyager-1 in deep space, i.e. beyond the heliopause. The density of ionizing radiation 
is about 3.21 times superior to the radiation dose observed nearby Earth. This nearby Earth radiation rate can be evaluated thanks to 
the Lunar Lander Neutrons and Dosimetry (LND) experiment aboard the Chinese Chang'E 4 mission that landed in the von Karman crater 
on the far side of the Moon on January 3, 2019. The LND measured a dose equivalent rate of 57.1 $\pm$ 10.6 $\mu$Sv per hour from 
charged particles \cite{Quan2020}. This converts into 447.6 $\pm$ 92.9 mSv per year. 

\begin{figure*}
\centering
  \subfloat[Evolution of the population within the space ship.]{\includegraphics[width=8cm]{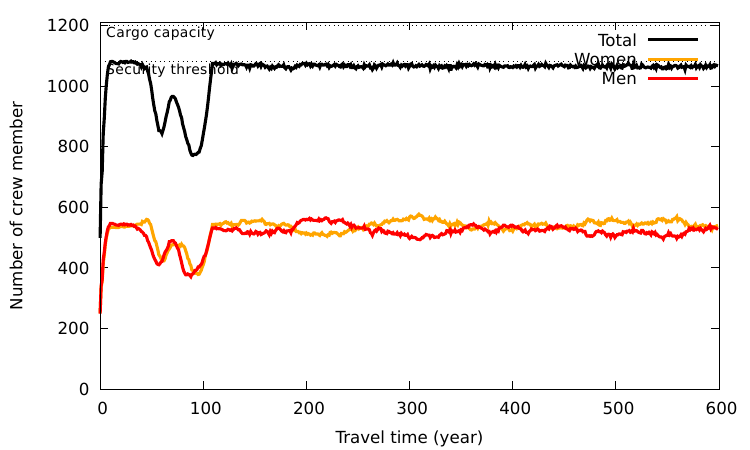}}\hfil
  \subfloat[Equivalent dose of ionizing radiation (in mSv).]{\includegraphics[width=8cm]{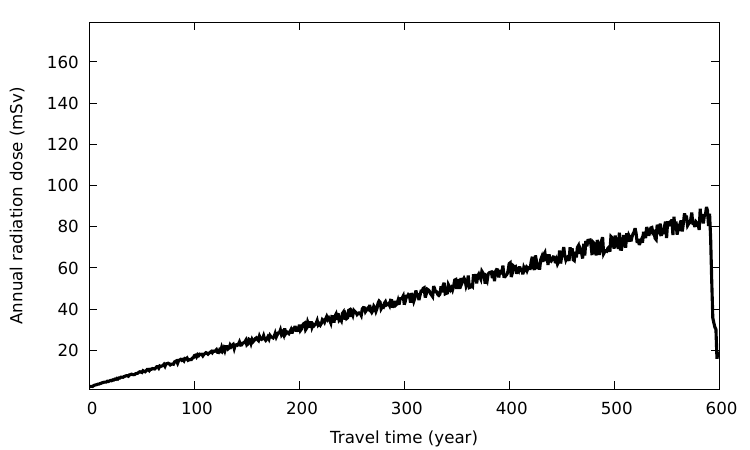}}

  \subfloat[Likelihood of infertility among breeding people.]{\includegraphics[width=8cm]{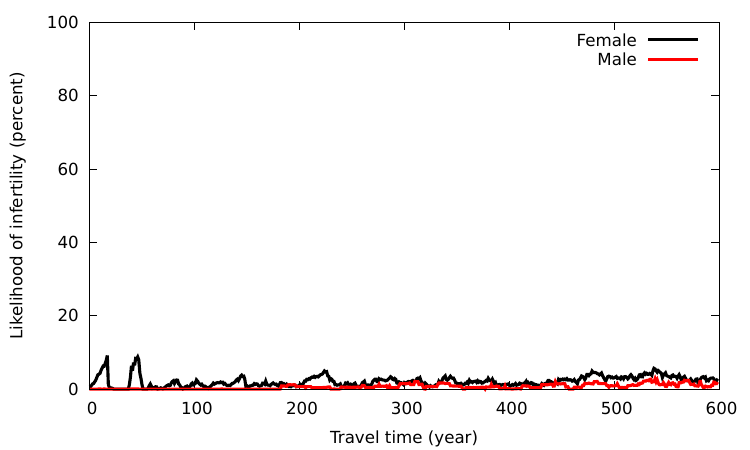}}\hfil
  \subfloat[Likelihood of pregnancy among breeding women.]{\includegraphics[width=8cm]{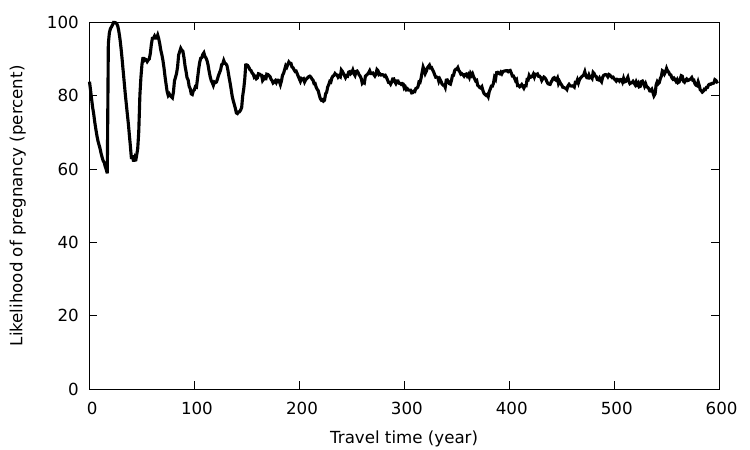}} 

  \subfloat[Number of miscarriages.]{\includegraphics[width=8cm]{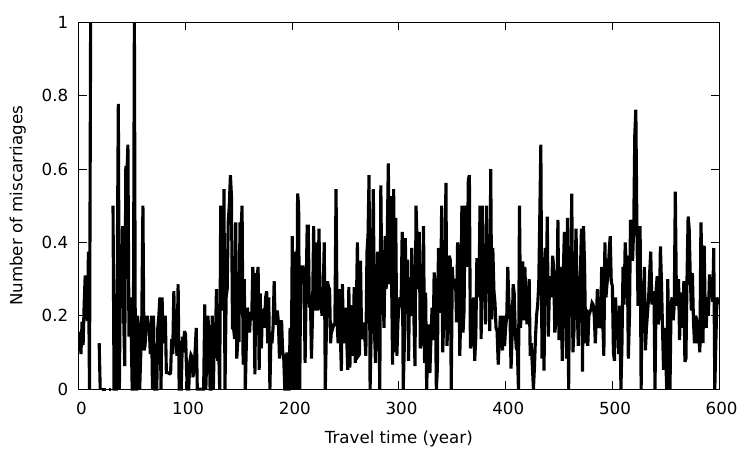}}\hfil   
  \subfloat[Individual heterozygosity among the crew.]{\includegraphics[width=8cm]{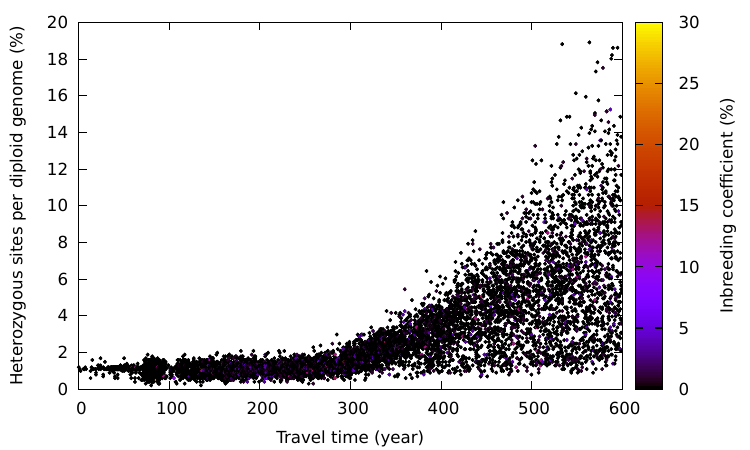}}
  \caption{Results from the third scenario (progressive degradation of the radiation shield).}
  \label{Fig:Scenario3}
\end{figure*}

We use this value as the radiation dose 
nearby Earth and scale this value according to the Voyager-1 count rate as a function of distance from our planet. It means that the 
radiation dose rate will slowly increase as we reach the heliopause, then plateau (within the measured radiation rate uncertainties) 
in the deep interstellar space between our solar system and the stellar system of the target exoplanet, where we postulate that a similar 
(but slightly more extended) heliopause exists, protecting the exoworld from the interstellar cosmic rays. The results of our simulations 
are shown in Fig.~\ref{Fig:Scenario3} and will be analyzed in details in Sect.~\ref{Analysis}.

\subsection{Scenario 4: nuclear incident}
\label{Results:nuclear}
Our fourth model investigates a Chernobyl-like catastrophe happening in the spacecraft. While the propulsion energy and technology at the time of 
future generation ships are unknown, we can postulate on the basis of current technology that nuclear reactors might be used \cite{Dyson2003}.
If an incident happens within the nuclear core of one of the engines, an explosion followed by contamination from radioactive particles will 
drastically increase the radiation levels aboard. The decay of the radioactive particle will take time and a confined fraction of the spacecraft
might be out of service for a while. Air would be contaminated too and carry away radiation, such as observed around Chernobyl \cite{Jaworowski2010}.

\begin{figure*}
\centering
  \subfloat[Evolution of the population within the space ship.]{\includegraphics[width=8cm]{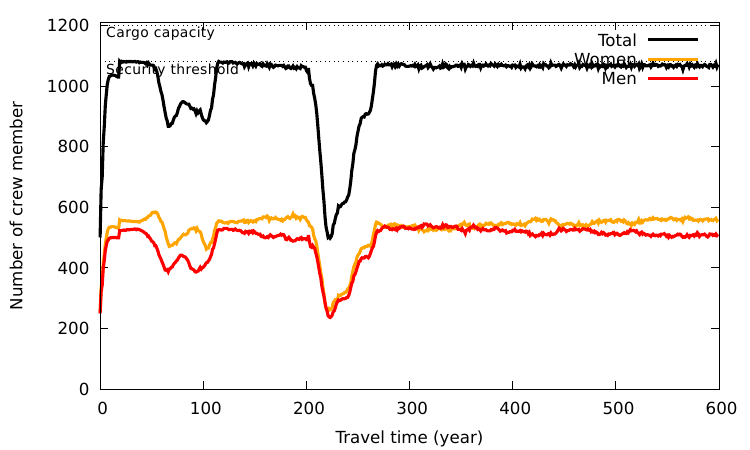}}\hfil
  \subfloat[Equivalent dose of ionizing radiation (in mSv).]{\includegraphics[width=8cm]{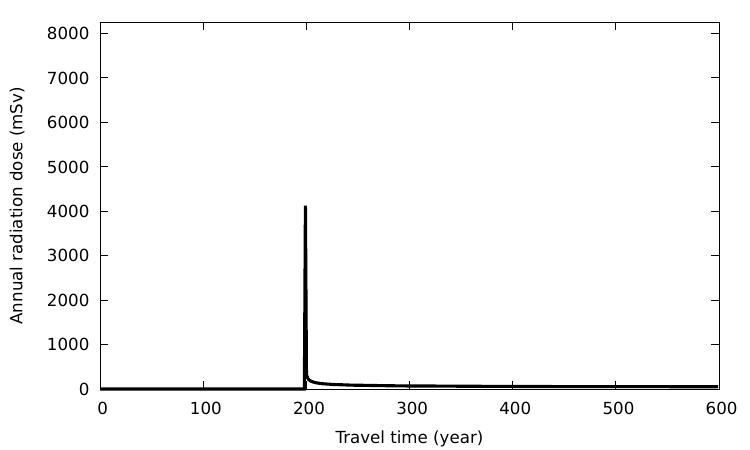}}

  \subfloat[Likelihood of infertility among breeding people.]{\includegraphics[width=8cm]{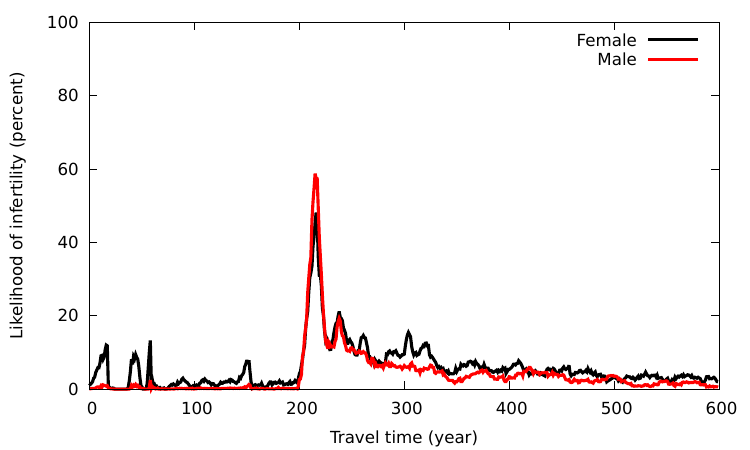}}\hfil
  \subfloat[Likelihood of pregnancy among breeding women.]{\includegraphics[width=8cm]{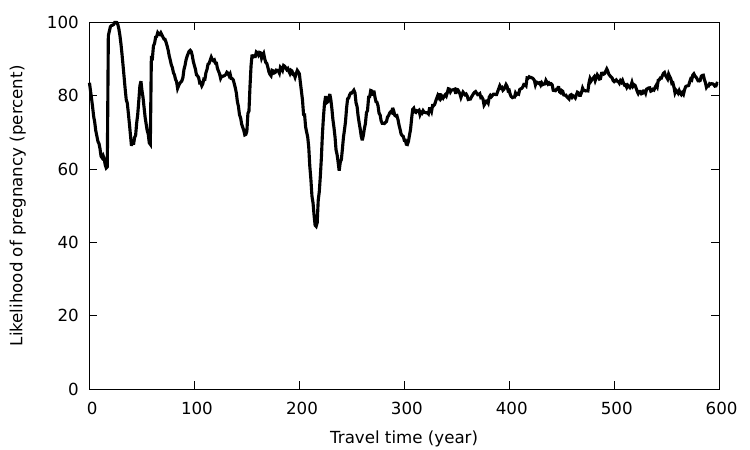}} 

  \subfloat[Number of miscarriages.]{\includegraphics[width=8cm]{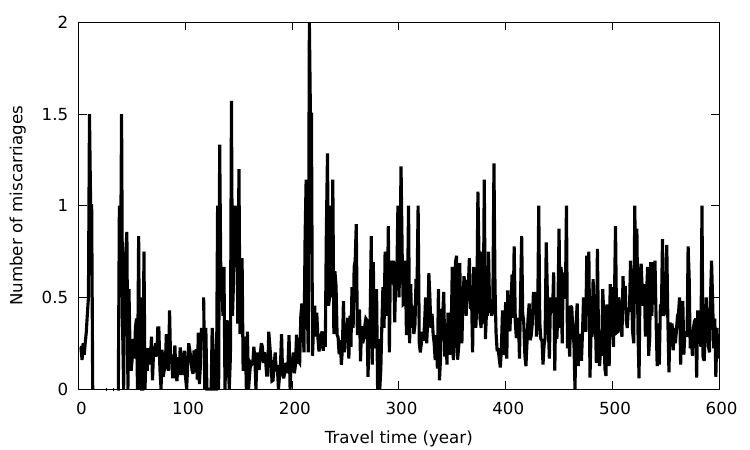}}\hfil   
  \subfloat[Individual heterozygosity among the crew.]{\includegraphics[width=8cm]{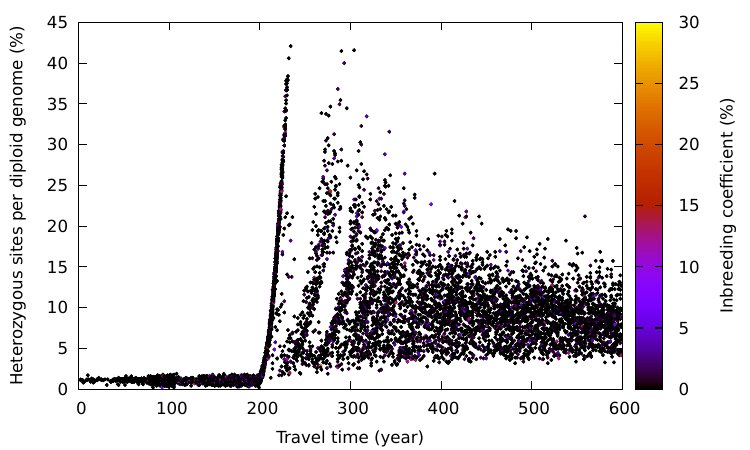}}
  \caption{Results from the fourth scenario (nuclear incident).}
  \label{Fig:Scenario4}
\end{figure*}

To mimic this effect, we retrieved the time-dependent evolution of the radiation dose-rate in the open air around Chernobyl \cite{Jonsson2017} and 
used a four parameter logistic (4PL) curve to fit the data (root-mean-square of the residuals: 0.7669). To estimate the initial dose rate, we 
compiled the biological dosimetric data from the Chernobyl accident. Based on dosimetry from a group of 134 hospitalized workers, the received 
whole-body doses from external irradiation averaged around 4.12~Sv \cite{Chernobyl2003}. While being a crude approximation, we use this number 
as the radiation rate for the first year and apply the 4PL curve to quantify the decrease of radiation rates. We set up this nuclear incident 
at year 200 after departure. The results of our simulations are shown in Fig.~\ref{Fig:Scenario4} and will be analyzed in details in Sect.~\ref{Analysis}.

\subsection{Scenario 5: supernovae}
\label{Results:supernovae}

Massive stars ($\ge$ 8 solar masses \cite{Heger2003}) undergo a series of stages before culminating in a supernova explosion. Beginning with nuclear fusion in their cores, stars progress through phases of hydrogen and helium burning until they exhaust their nuclear fuel. As the core contracts and heats up, fusion reactions continue in surrounding shells, creating heavier elements until iron accumulates in the core. The sudden collapse of the iron core triggers a shockwave that leads to a violent explosion, ejecting the star's outer layers into space and producing a supernova. These explosions play a crucial role in the universe's evolution, dispersing heavy elements and radiation. 

Exoplanets close to the supernova would be bombarded with intense radiation, including $\gamma$-rays, X-rays and ultraviolet radiation. Those radiation could strip away their atmospheres, ionize surface materials and cause biological damage to any potential life forms. A similar destiny would await any spaceship traveling too close to a supernova, so it is an interesting (yet scholar) hypothesis to test using our code. In this subsection, we thus study the impact such an explosion (and its associated intense radiation field) could have on the crew of a generation ship. To do so, we need to determine how much radiation is ejected from a stellar explosion : we first compute the injected spectrum of a supernova and then its diffused counterpart.

\subsubsection{Injected spectrum}
In the context of supernovae, the injected spectrum refers to the release (or injection) of cosmic rays into the surrounding space. The mathematical form of the injected spectrum is often characterized by a power-law distribution, expressed as:
\begin{equation}
    Q_{\rm{CR}}(E) = A \left(\frac{E}{E_0}\right)^{-\alpha}.
\end{equation}
Here, $Q_{\rm{CR}}(E)$ represents the cosmic ray spectrum in units of erg$^{-1}$ as a function of energy $E$, $A$ is the amplitude determining the overall normalization, $E_0$ is a characteristic energy scale and $\alpha$ is the spectral index. The $\alpha$ index typically hovers around 2.7 and the choice of $E_0$ influences the shape of the spectrum. The amplitude $A$ is intricately linked to the fraction of explosion energy from the supernova ($\xi_{\rm{CR}}$) and the energy bounds ($E_{\rm{min}}$ to $E_{\rm{max}}$). For the remaining of the computation, we will consider that a fraction $\xi_{\rm CR} \sim 0.1$ of the total explosion energy $E_{\rm SN} \sim 10^{51}$ erg is converted into cosmic rays, so that:
\begin{equation}
    \int_{E_{\rm min}}^{E_{\rm max}}Q_{\rm CR}(E) E {\rm d}E = \xi_{\rm CR} E_{\rm SN},
\end{equation}
which leads to:
\begin{equation}
    A= \frac{\xi_{\rm CR}  E_{\rm SN}}{(\alpha-2) E_{0}^{\alpha} \left( \frac{1}{E_{\rm min}^{(\alpha-2)}} -\frac{1}{E_{\rm max}^{(\alpha-2)}} \right) }.
\end{equation}
Therefore, if $E_{\rm max}$ is sufficiently large and $E_{\rm min}\sim E_0 \sim 1$ GeV, then:
\begin{equation}
A= \frac{\xi_{\rm CR} E_{\rm SN}}{(\alpha-2)E_0^2},
\end{equation}
resulting in:
\begin{equation}
Q_{\rm CR}(E) \sim \frac{\xi_{\rm CR} E_{\rm SN}}{(\alpha-2)E_0^2} \left( \frac{E}{E_0} \right)^{- \alpha}.
\end{equation}

In essence, the injected spectrum provides a quantitative description of the abundance of cosmic rays across different energy levels resulting from the astrophysical processes associated with supernovae.

\subsubsection{Diffused spectrum}
The diffused spectrum refers to the distribution of cosmic rays that have propagated and dispersed through space following their injection into the cosmic environment. Unlike the injected spectrum, which characterizes the immediate release of cosmic rays from their source, the diffused spectrum accounts for the spread and diffusion of these particles over time and distance.

The spectrum \(Q_{\rm{CR}}(E)\) is injected at an instant \(t_0\) at a position \(\vec{r_0}\). To determine the density of cosmic rays at different energy levels and positions \(\vec{r}\) in space as they evolve over time \(t\), we can solve the diffusion equation using Green's formalism:

\begin{equation}
\frac{\partial G}{\partial t} - D(E) \vec{\nabla^2} G = Q_{\rm CR}(E) \delta(\vec{r}-\vec{r_0}) \delta(t-t_0).
\end{equation}

This equation describes how cosmic rays disperse in space as a function of time, considering the interplay between various factors such as the energy of the particles, the surrounding medium's properties and the initial conditions set by the astrophysical events. The solution to this equation can be expressed as:

\begin{equation}
    \begin{split}
        G(t,\vec{r},E, t_0, \vec{r_0}) &= \frac{Q_{\rm CR}(E)}{\left[4 \pi D(E)(t-t_0)\right]^{3/2}} \\
        &\times \exp \left[- \frac{(\vec{r}-\vec{r_0})^2}{4 D(E)(t-t_0)}\right]
    \end{split}
\end{equation}

where \(G\) is in units of [E$^{-1}$ cm$^{-3}$]. The diffusion coefficient \(D(E)\) is not well known, but a first-order approximation can be given by:

\begin{equation}
D(E) \approx 10^{28} \left( \frac{E}{10 {\rm GeV}}\right)^{\delta}
\end{equation}

with \(\delta \sim 0.3-0.7\). For the remainder of this paper, we will use a medium value \(\delta = 0.5\).

\subsubsection{Conversion to radiation dose}
Once the supernova-enriched density of cosmic rays at a given position in time and space is known, we have to convert this information into unshielded absorption doses ($D$). To do so, we rely on the estimation provided by \cite{Semyonov2006} :

\begin{equation}
D = 0.53 \times N \times H \times d \times S / M
\end{equation}

where $N$ is the number of nucleons per square centimeter per second from all directions through the ship’s cross-section perpendicular to the direction of cruising, $H$ is the stopping power of 1 GeV cosmic rays in tissue (or in water, in MeV per centimeter), $d$ is the average thickness of the human torso in centimeters, $S$ is the average cross-section of the human body in square centimeter and $M$ is the approximate mass of an adult astronaut in grams. This gives us the unshielded absorption doses in rads per year that we can convert to the equivalent doses of unshielded radiation by multiplying $D$ with the radiation quality factor $Q$ ($Q$ = 10 for neutrons and high energy protons according to NRC Regulations).

\subsubsection{A supernova 50 light-years away}
As seen in the previous subsections, the amount of cosmic rays originating from a supernova strongly depends on the distance between the cataclysmic event and the observer (the interstellar spaceship here), as well as the time since the explosion. To test the impact of a supernova onto the radiation levels measured at the position of the generation ship, we situate the exploding star at a distance of 50 light-years.  

\begin{figure*}
\centering
  \subfloat[Evolution of the population within the space ship.]{\includegraphics[width=8cm]{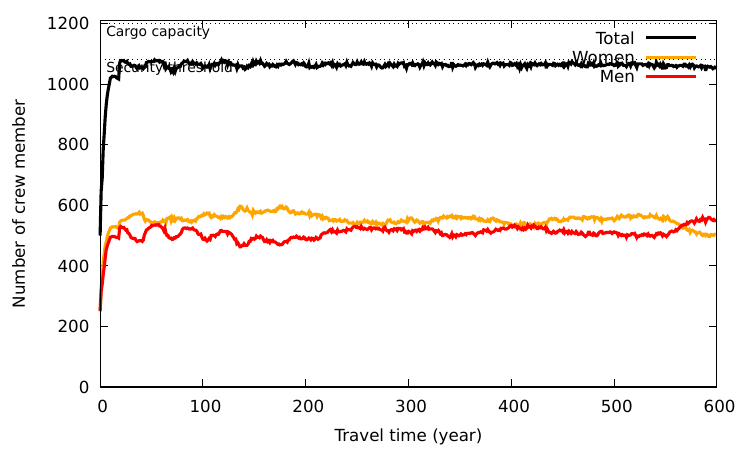}}\hfil
  \subfloat[Equivalent dose of ionizing radiation (in mSv).]{\includegraphics[width=8cm]{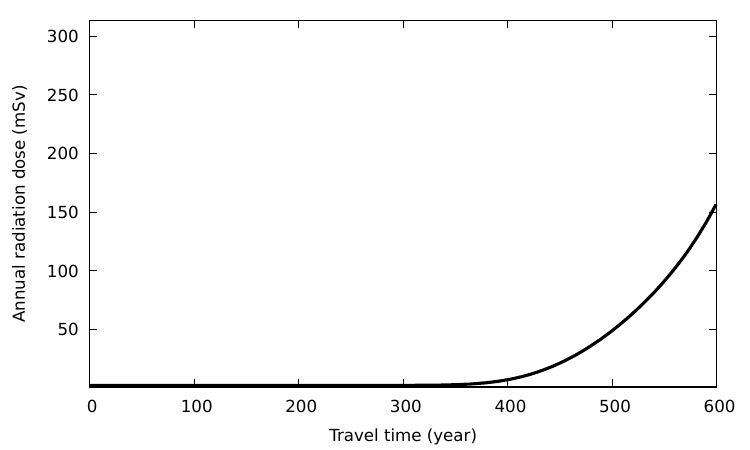}}

  \subfloat[Likelihood of infertility among breeding people.]{\includegraphics[width=8cm]{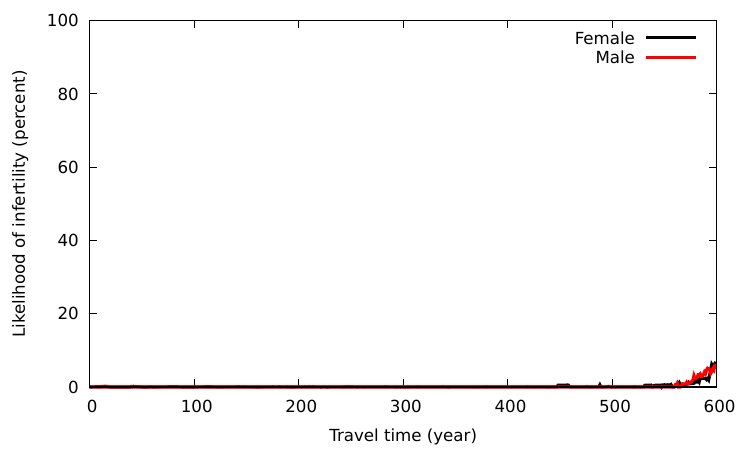}}\hfil
  \subfloat[Likelihood of pregnancy among breeding women.]{\includegraphics[width=8cm]{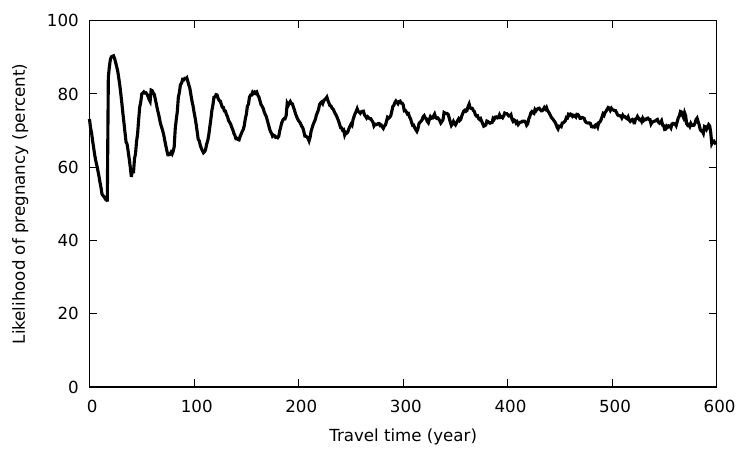}} 

  \subfloat[Number of miscarriages.]{\includegraphics[width=8cm]{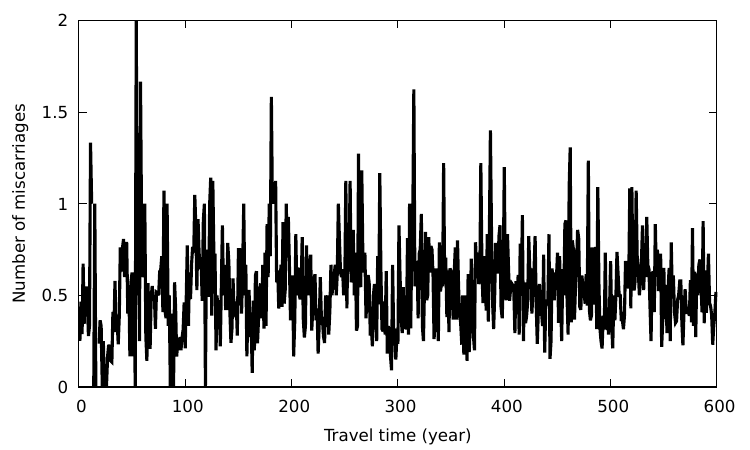}}\hfil   
  \subfloat[Individual heterozygosity among the crew.]{\includegraphics[width=8cm]{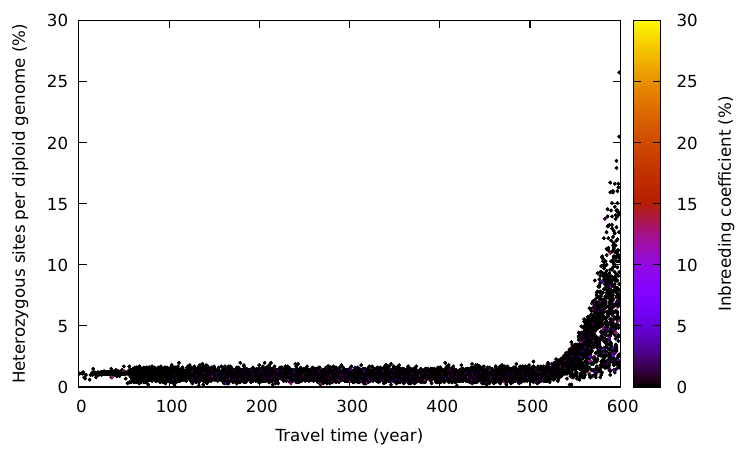}}
  \caption{Results from the fifth scenario (supernova situated at 50 light-years).}
  \label{Fig:Scenario5}
\end{figure*}

Such distance was not selected at random. Recent measurements of iron 60 ($^{60}$Fe) in the Earth seabed, an isotope of iron produced in supernova explosions found in fossilized bacteria in sediments on the ocean floor, indicate that the Earth suffered from, at least, two supernovae several million years ago \cite{Melott2017}. The estimated distances of those supernovae ranges from $\sim$ 160 to $\sim$ 325 light-years, indicating that the kill-zone of supernovae (the distance between the supernova and the Earth that would trigger a mass extinction) is much smaller. The first detailed computation of the kill-zone estimated a kill-zone radius of about 25 light-years \cite{Gehrels2003}, but current simulations tend to revise these figures upwards. A more probable radius for the kill-zone is 35 - 65 light-years \cite{Thomas2023,Perkins2024}. We thus opted for a median value of 50 light-years. Similarly to the previous scenario, we set up this stellar explosion at year 200 after departure. The results of our simulations are shown in Fig.~\ref{Fig:Scenario5} and will be analyzed in details in Sect.~\ref{Analysis}.

%%%%%%%%%%%%%%%%%%%%%%%%%%%%%%%%%%%%%%%%%%%%%%%%%%%%%%%%%%%%%%%%%%%%%%%%%%%%
%%%%%%%%%%%%%%%%%%%%%%%%%%%%%%%%%%%%%%%%%%%%%%%%%%%%%%%%%%%%%%%%%%%%%%%%%%%%
\section{Analysis of our simulations}
\label{Analysis}

We have produced a set of five simulations to explore the impact of various radiation situations that could happen during a deep space travel. We now examine each of them separately. Tab.~\ref{Tab:Results} compiles several key quantitative indicators of the genetic health of the crew. Among them are the initial and final genome diversities with respect to the reference genome presented in Sect.~\ref{Modeling}, the initial and final degrees of genetic polymorphism, the initial and final degrees of genetic heterozygosity and the Nei's genetic distance (as described in \cite{Chakraborty2011}). Haplotype heat maps of all chromosomes from the initial and final populations are not shown but are available upon reasonable request.

\subsection{Terrestrial background}
In the case of the first simulation (``terrestrial background'' scenario, see Fig.~\ref{Fig:Scenario1}), the crew swiftly reaches the maximum generation ship capacity (set by the security threshold) and the demography then stagnates for the rest of the travel at about 1\,080 people per year in the spaceship. The numbers of women and men are approximatively the same over the whole trip, with a slightly higher but non-statistically significant number of women. The annual radiation dose, set at an averaged terrestrial background level (2.4~mSv per year), has no impact on the likelihood of infertility, miscarriage or pregnancy chances. 

The values are quite similar to the one found in the first papers of our HERITAGE series \cite{Marin2017,Marin2018,Marin2020}. The individual heterozygosity among the crew is stable in time. There is no differences in initial and final genome diversities, genetic polymorphism and genetic heterozygosity. The computed Nei's genetic distance indicates that the initial and final populations are very similar (in the sense that they share very similar allelic patterns) and likely poorly differentiated. 

As a consequence, this scenario demonstrates that phenotypical effects are close to the neutral hypothesis as the frequency of alleles (for non-sexual chromosomes) is stable over centuries for large departing crews (500 people or more).

\subsection{Extreme background}
Looking at the same scenario but with higher radiation background levels (``extreme background'' scenario, see Fig.~\ref{Fig:Scenario2}), we see that the crew seems to be able to quickly reach the same population demography as in the first scenario. However, in comparison to the previous simulation, twice the number of crew members were created. This is due to the fact that the high radiation levels have a negative impact onto the life expectancy of the crew : neo-mutations make the infertility and miscarriage rates rise, while they diminish the chances of getting pregnant with time. This is a clear signature of the phenotypic effects of gene expression. 

This is also illustrated in the time-evolution of individual heterozygosity among the crew. The baseline level increases with time, as more neo-mutations are transmitted from a generation to the new one. The spread in heterozygosity is constant with time (about 30 percentage points). The impact of neo-mutations can also be seen in the genome diversities that reaches 8.97\% at the end of the simulation, while it was fixed at 0.57\% at the beginning. The degree of polymorphism, that represents the fraction of loci (among the total N loci) that are polymorphic at the population scale started at 9.00\% and reached 100\% by the end of the simulation. The computed Nei's genetic distance reaches 0.23\%, a value that is big enough to arbitrarily subdivide the initial and final crews into distinct entities (subspecies, such as, e.g. the Bengal and Siberian tigers). 

The constant rise in time of infertility among crew members, associated with the declining pregnancy chances, indicate that such population is likely to die within the next centuries/millennium without the contribution of a new genetic pool or the drastic reduction of the radiation levels.

\subsection{Shield degradation}
The third scenario implies a time-dependent degradation of the ship protection against radiation (``shield degradation'' scenario, see Fig.~\ref{Fig:Scenario3}). Similarly to the first scenario (``terrestrial background''), the generation ship is quickly inhabited at its fullest and the infertility and miscarriage rates are very low during most of the travel time. The averaged chances of getting pregnant remain high too. 

However, with the degradation of the shields with time, we observe a slight tendency of the aforementioned indicators to vary by the end of the simulation. This explains why 9\,519 crew members were simulated, instead of the 7\,015 of the first simulation. More importantly, the averaged individual heterozygosity among the crew rises with time, as well as its spread (from a percentage point at the beginning to about 20 percentage points at the end of the simulation). Genome diversities and genetic polymorphism are also on the rise, with a crew showing 100\% polymorphism by the end of the simulation. However, the Nei's genetic distance remains low, similar to the first scenario, since the genome diversity only gained one percentage point. 

We note that the final background level inside the generation ship is still below the extreme background value of the second simulation, despite a 0.01\% effectiveness degradation per year, explaining why the final crew is in better genetic health than in a scenario where they constantly endure large radiation doses.

\subsection{Nuclear incident}
In case of a Chernobyl-like incident inside the generation ship (``Nuclear incident'' scenario, see Fig.~\ref{Fig:Scenario4}), the first 200 years of the mission are very similar to the first case, as the input parameters are the same. But when the nuclear incident onsets, there is a quick rise of the radiation levels, leading to deadly cancers and many malformation and miscarriages that strongly and negatively impact the crew. The population was almost extinguished by the event. Despite the incident happening only at a single time step (one year), the crew took about 75 years to recover in terms of population number. 

The nuclear incident is linked with a sudden increase in infertility and miscarriage rates that keeps echoing until the end of the simulation. It also affected, but less severely, the chances of being pregnant. The echoes of the nuclear incident are clearly seen in the graph showing the individual heterozygosity among the crew. At the time of the nuclear catastrophe, the heterozygosity rose by 40 percentage points but, as time passes and the fallout settles and loses radiative power, the heterozygosity shows a decreasing trend. At the end, the genome diversitiy (2.8\%) is higher than in the first and third cases and polymorphism is maximum (100\%), leading to a slightly larger Nei's genetic distance (0.06\%) than in a smiluation without a nuclear incident. This genetic distance value is insufficient to unreservedly differentiate the initial and final crews in terms of genetics.

\subsection{Supernova}
Finally, the stellar explosion case (``Supernova (50~ly away)'' scenario, see Fig.~\ref{Fig:Scenario5}) shows that the crew is completely safe during the first two-third of the mission. Despite having exploded 200 years after departure, the cosmic rays injected in space were spread and diffused by interstellar particles over time and distance, explaining why the levels of radiation only start to rise at the middle of the mission. 

The annual dose of radiation increases rapidly with time, impacting the likelihood of infertility, miscarriage and pregnancy chances only before the end of the simulation. This is clearly visible in the graph showing the individual heterozygosity among the crew that presents an exponential behavior during the last 50 years of the simulation. This leads to a subtle increase of the genome diversity, genetic polymorphism and heterozygosity but the supernova did not have time to strongly alter the human genome, so the Nei's genetic distance remains negligible. 

The simulation stops at year 600 but, in the future, it will be interesting to increase this duration to determine how far a supernova must be to fully annihilate a population in a spacecraft or on Earth. This will be explore in another publication.

\begin{table*}[h!]
  \begin{center}
    \begin{tabular}{l|c|c|c|c|c|c|c|c}
      \textbf{Scenario} & \textbf{N$_{\rm tot}$} & \textbf{G$_{\rm i}$} & \textbf{G$_{\rm f}$} & \textbf{P$_{\rm i}$} & \textbf{P$_{\rm f}$} & \textbf{H$_{\rm i}$} & \textbf{H$_{\rm f}$} & \textbf{D$_{\rm Nei}$}\\ 
      \hline
      Terrestrial background & 7\,015 & 0.57\% & 0.56\% & 9.31\% & 9.70\% & 1.13\% & 0.98\% & 0.02\% \\
      Extreme background & 15\,378 & 0.57\% & 8.97\% & 9.00\% & 100\% & 1.13\% & 15.01\% & 0.23\% \\
      Shield degradation & 9\,519 & 0.57\% & 1.58\% & 9.70\% & 100\% & 1.12\% & 2.29\% & 0.02\% \\
      Nuclear incident & 9\,190 & 0.56\% & 2.8\% & 9.13\% & 100\% & 1.16\% & 4.98\% & 0.06\% \\
      Supernova (50~ly away) & 7\,666 & 0.56\% & 0.59\% & 8.22\% & 9.11\% & 1.11\% & 1.03\% & 0.01\% \\
    \end{tabular}
    \caption{Results from the five scenarios investigated in this paper. 
	     N$_{\rm tot}$ designates the total number of digital humans 
	     that lived during the 600 years simulation. G$_{\rm i}$ and 
	     G$_{\rm f}$ are the initial and final genome diversities 
	     with respect to the reference genome, P$_{\rm i}$ and 
	     P$_{\rm f}$ are the initial and final degrees of genetic 
	     polymorphism, and H$_{\rm i}$ and H$_{\rm f}$ are the initial 
	     and final degrees of genetic heterozygosity. D$_{\rm Nei}$
	     is the Nei's genetic distance, as described in \cite{Chakraborty2011}.}
    \label{Tab:Results}
  \end{center}
\end{table*}

%%%%%%%%%%%%%%%%%%%%%%%%%%%%%%%%%%%%%%%%%%%%%%%%%%%%%%%%%%%%%%%%%%%%%%%%%%%%
%%%%%%%%%%%%%%%%%%%%%%%%%%%%%%%%%%%%%%%%%%%%%%%%%%%%%%%%%%%%%%%%%%%%%%%%%%%%
\section{Conclusions and further development}
\label{Conclusions}

We have upgraded the HERITAGE code in order to include genetic effects of mutation and neo-mutations (from radiations) onto the population's life expectancy, fertility, pregnancy chances and miscarriage rates. The code is modular enough so that the parametrization of the genetic effects of mutation and neo-mutations can be changed according to the user's needs. Using a Gaussian function for the mutations and a bimodal distribution for the effects of neo-mutations, we have shown the impact of phenotypic effects of gene expression onto a population of digital humans aboard a generation ship towards a distant exoplanet. Over a 600 years-long travel, we tested different radiation levels and incidents that could realistically happen, in order to check the resilience of the crew and the effects of mutations. 

Our code can now be used to simulate the immediate or progressive degradation of an environment enduring strong radiation fields in order to assess the hazard levels and viability of humans groups under those conditions. HERITAGE can be of great interest to test how the degradation in time of the protective shields abord space shuttles and space station can affect the crew, allowing better decision-making in the duration of manned missions. Also, it can be of theoretical use to determine the minimum distance between a supernova and the spaceship for the interstellar ark to survive, permitting us to effectively determine the safest space route to reach an exoplanet. Most of the current simulations focus on the impact of supernova onto the biota of the Earth but not on a human population in space, where the Earth's magnetic shield has to be replaced with articial shields (such as hydrogen rich protective layers \cite{Cha2022}). 

The code is at a level where most of the critical parameters for testing the survivability of a crew in space have beend added, even though the code could still be upgraded to include subtle population dynamics effects. HERITAGE will now be used to test various scenarios to optimize the crew selection (the ideal departing crew number, the usefulness of gender balance, the use of frozen genetic material...), to test generation ship designs, propulsions and radiation shields, and to pave the way for building the first generation of space arks.

%%%%%%%%%%%%%%%%%%%%%%%%%%%%%%%%%%%%%%%%%%%%%%%%%%%%%%%%%%%%%%%%%%%%%%%%%%%%
%%%%%%%%%%%%%%%%%%%%%%%%%%%%%%%%%%%%%%%%%%%%%%%%%%%%%%%%%%%%%%%%%%%%%%%%%%%%
\section*{Acknowledgment}
The authors would like to acknowledge Dr. Pierre Cristofari for his help in computing the diffused spectrum of a supernova explosion. 
We are also gratful to Dr. Rhys Taylor for his comments and suggestions that greatly helped to improve this paper.

\bibliographystyle{elsarticle-num}
\bibliography{mybibfile}

\end{document}